\documentclass{PoS}

\title{Chiral Thermodynamics with Charm}

\ShortTitle{Chiral thermodynamics with charm}

\author{\speaker{Chihiro Sasaki}\\
        Frankfurt Institute for Advanced Studies,
        D-60438 Frankfurt am Main, Germany\\
        Institute of Theoretical Physics, University of Wroclaw,
        PL-50204 Wroclaw, Poland\\
        E-mail: \email{sasaki@fias.uni-frankfurt.de}}

\abstract{
Chiral thermodynamics of charmed mesons is formulated 
at finite temperature within a $2+1+1$-flavored effective 
Lagrangian incorporating heavy quark symmetry. 
The chiral mass splittings are shown to be less sensitive 
to the light-quark flavors, attributed to the 
underlying heavy quark symmetry. Consequently, chiral symmetry 
restoration is more accelerated in the strange charmed-mesons
 than in the strange light mesons, and this is in striking 
contrast to the chiral SU(4) result. We also show the correlations
between light and heavy flavors around the chiral crossover.
}

\FullConference{9th International Workshop on Critical Point and Onset of Deconfinement\\
		 17-21 November, 2014\\
		 ZiF (Center of Interdisciplinary Research), University of Bielefeld, Germany}

\begin{document}

\section{Introduction}

Heavy flavors are produced at the initial stage of
the high-energy heavy-ion collisions, so that they are expected to carry
the dynamical history of a created matter, the Quark-Gluon Plasma (QGP). 
Recent experimental observations have
revealed that charm quarks are thermalized~\cite{phenix,star,alice1,alice2},
contrary to earlier anticipation. Charge fluctuations calculated in lattice 
QCD also indicate that the charmed mesons are deconfined together with 
light-flavor mesons~\cite{latcharm}. Given those observations, comprehensive
exploration for the chiral aspects of the heavy-light hadrons increases its
importance.

In constructing effective Lagrangians for the heavy-light mesons, 
besides spontaneous chiral symmetry breaking, heavy quark symmetry is 
a vital ingredient~\cite{HQS}.
The pseudo-scalar $D$ and vector $D^\ast$ states fill in the same 
multiplet $H$, forming the lowest spin partners. Their low-energy 
dynamics is dominated by interactions with Nambu-Goldstone (NG) bosons, 
pions~\cite{Georgi,Wise,BD,YCCLLY}.
Introducing the multiplet including $D$ and $D^\ast$ inevitably 
accompanies another multiplet $G$ which contains a scalar $D_0^\ast$ 
and axial-vector $D_1$ states. Those parity partners, $H$ and $G$, 
become degenerate when the chiral symmetry is restored~\cite{NRZ,BH}.

Aside from the chiral $SU(4)$ approach where the charm sector suffers 
from a huge explicit breaking of the extended flavor symmetry, 
a self-consistent study for the thermal charmed-mesons with implementing 
heavy quark symmetry has received little attention. In Ref.~\cite{charm:cs},
a chiral effective theory for the light and heavy-light mesons has been
formulated in the presence of a medium. Below, we will briefly discuss
in-medium masses of the heavy-light mesons and the correlations among 
light and heavy flavors near the chiral crossover.

\section{Chiral mass splittings}

When the charmed-meson mean fields are introduced to a chiral effective
theory in the standard fashion, they act as an extra source which breaks
the chiral symmetry explicitly. Consequently, an unrealistically strong 
mixing between the light-flavor and the charmed meson sector is induced.
This defect can be avoided if effective interactions depending on temperature
are introduced. Those intrinsic modifications can be extracted from the 
chiral condensates calculated in lattice QCD. The obtained coupling of 
the strange charmed meson to the sigma meson, $g_\pi^s(T)$, becomes quenched 
as temperature is increased toward the chiral pseudo-critical point 
$T_{\rm pc} = 154$ MeV.

\begin{figure}
\begin{center}
\includegraphics[width=7cm]{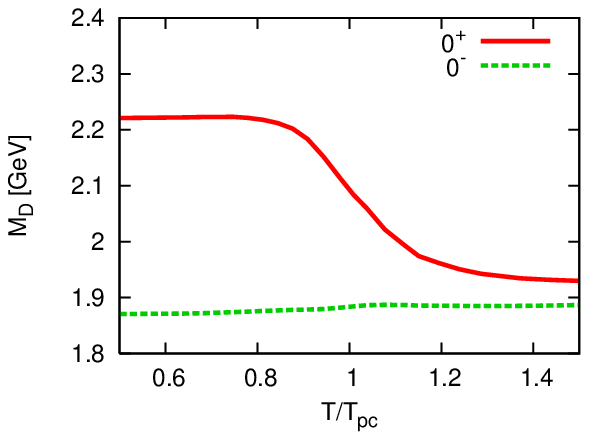}
\includegraphics[width=7cm]{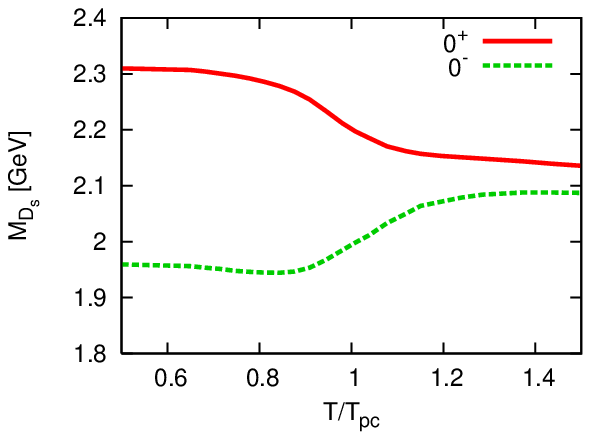}
\caption{
In-medium masses of the non-strange (left) and strange (right)
charmed-mesons with positive and negative parity~\cite{charm:cs}.
}
\label{fig:md}
\end{center}
\end{figure}

Effective charmed-meson masses in hot matter are shown in Fig.~\ref{fig:md}.
The parity partners approach each other as temperature is increased
both in the non-strange and strange sector, in consistent with the chiral
restoration. The two pseudo-scalar states have the same trend that their
masses are increasing with temperature, although the non-strange meson mass
exhibits a rather weak modification. On the other hand, the two scalar states
drop significantly; the non-strange meson mass by $\sim 200$ MeV and 
the strange meson mass by $\sim 100$ MeV. The mass splittings between 
the non-strange and strange states are around $200$ MeV above $T_{\rm pc}$ 
due to the fact that the chiral symmetry in the strange sector is not 
restored yet.
Nevertheless, the chiral mass splittings between the scalar and 
pseudo-scalar states are almost of the same size,
\begin{equation}
\delta M_D(T_{\rm pc}) \sim \delta M_{D_s}(T_{\rm pc}) 
\sim 200\,\mbox{MeV}\,,
\label{HLchiral}
\end{equation}
i.e. {\it the chiral mass differences in the heavy-light sector are 
blind to the light flavors.} This is a striking difference from the 
chiral properties of the light mesons, and is attributed to the heavy 
quark symmetry possessed by the leading-order Lagrangian in $1/m_Q$ 
expansion. In contrast, the chiral SU(4) model, where the charmed mesons 
are treated on the equal footing to the non-strange and strange mesons, 
yields a qualitatively different result from Eq.~(\ref{HLchiral}); 
$\delta M_D$ is much smaller than 
$\delta M_{D_s}$, similar to the light meson masses~\cite{su4}.

The quenched $g_\pi^s(T)$ leads also to a strong suppression of the 
scalar $D_s$ decay toward $T_{\rm pc}$, on top of the suppression due 
to the small isospin violation. Thus, such an anomalous suppression,
if it would be observed, is a signature of chiral symmetry restoration. 
The same should be carried over to 
the $B$ and $B_s$ mesons with which the heavy quark symmetry is 
more reliable.

\section{Flavor correlations}

Within the same theoretical framework,
thermal fluctuations and correlations between the light and heavy-light
mesons can also be studied. The extended susceptibility is given in the 
following matrix form~\cite{charm:fluc}:
\begin{eqnarray}
\hat{\chi}_{\sigma\sigma}
&=&
\hat{\chi}_{\rm ch} + \hat{\chi}_{\rm ch}\hat{\mathcal C}_{\rm HL}
\hat{\chi}_D\hat{\mathcal C}_{\rm HL}\hat{\chi}_{\rm ch}\,,
\nonumber\\
\hat{\chi}_{\sigma D}
&=&
-\hat{\chi}_{\rm ch}\hat{\mathcal C}_{\rm HL}\hat{\chi}_D\,,
\nonumber\\
\hat{\chi}_{D\sigma}
&=&
-\hat{\chi}_D\hat{\mathcal C}_{\rm HL}\hat{\chi}_{\rm ch}\,,
\nonumber\\
\hat{\chi}_{DD}
&=&
\hat{\mathcal C}_D - \hat{\mathcal C}_{\rm HL}\hat{\chi}_{\rm ch}
\hat{\mathcal C}_{\rm HL}
\equiv
\hat{\chi}_D\,,
\end{eqnarray}
where $\hat{\chi}_{\rm ch}$ is the chiral susceptibilities of non-strange
and strange quark condensates which are responsible for the critical
behaviors, and $\hat{\mathcal C}_{{\rm HL},D}$ represents the curvature 
of the effective potential.

\begin{figure}
\begin{center}
\includegraphics[width=7cm]{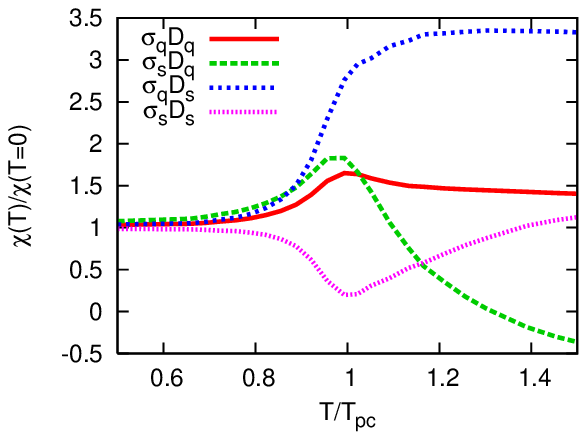}
\includegraphics[width=7cm]{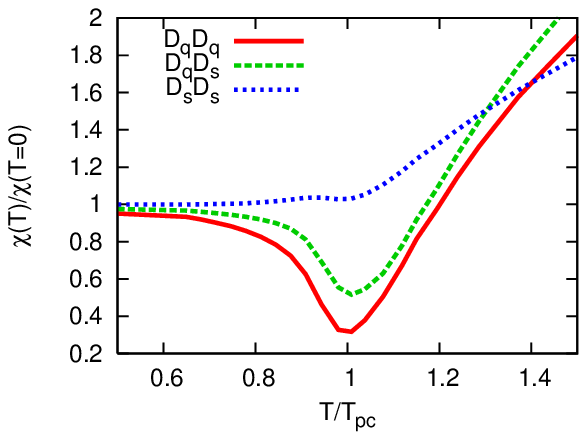}
\caption{
Flavor correlations between the light and heavy-light mesons (left)
and among heavy-light mesons (right).
}
\label{fig:susHL}
\end{center}
\end{figure}

Correlations between the light and heavy-light mesons, as well as  
those between the heavy-light mesons, are shown  in
Fig.~\ref{fig:susHL}.
Due to the embedding of $\hat{\chi}_{\rm ch}$, 
various heavy-light flavor correlations are modified qualitatively
in a narrow range of temperature where the chiral susceptibility exhibits
a peak.
The onset of the chiral crossover, in the heavy-light flavor correlations,   
is therefore independent from the
light flavors. This indicates that the fluctuations carried by strange
charmed mesons can also be used to identify the chiral crossover,   
which is  dominated by the non-strange light quark dynamics.

\section{Conclusions}

The chiral mass splittings are shown to be essentially insensitive 
to the light-quark flavors, in spite of a non-negligible explicit 
breaking of the chiral $SU(3)$ symmetry. 
This ``blindness'' of the charm quark to the light degrees of freedom is 
dictated by the heavy quark symmetry. In contrast,
the kaon and its chiral partner masses become degenerate at a higher 
temperature than $T_{\rm pc}$, indicating a delay of the $SU(3)$ symmetry 
restoration. In the heavy-light sector, on the other hand, the strange 
charmed meson captures the onset of chiral symmetry restoration more 
strongly than the strange light meson does. 
A similar consequence is found in the correlations involving the heavy-light
meson mean fields. Those fluctuations exhibit certain qualitative changes
around $T_{\rm pc}$, and this feature is independent of the light flavors.
Hence, the strange charmed mesons have a potential for measuring a remnant
of $O(4)$.

It is a real challenge to explore the thermodynamics of heavy-light 
hadrons at high density.
A central task is to introduce a reliable density dependence of the 
interaction parameters, which requires either a more microscopic
prescription in the effective theory side, or more precise data for 
the chiral condensates from lattice simulations.
There are some implications of a confined phase with unbroken 
chiral symmetry from the lattice studies where the low-lying Dirac
modes, which generate a non-vanishing chiral condensate via 
the Banks-Casher relation, are removed. This new state of hadronic 
matter might appear in the QCD phase diagram at high baryon density,
where chiral hadronic theories may become valid in a wider parameter
range (See Ref.~\cite{cs:qm} and references therein).

\subsection*{Acknowledgments}

The work has been partly supported by the Hessian LOEWE initiative
through the Helmholtz International Center for FAIR (HIC for FAIR),
and by the Polish Science Foundation (NCN) under
Maestro grant DEC-2013/10/A/ST2/00106.


\end{document}